\documentclass[aps,prd,twocolumn,10pts]{revtex4}
\usepackage{amsfonts,amssymb,amscd,amsmath}
\usepackage{graphicx}
\usepackage{epsfig}
\usepackage{latexsym}
\usepackage{amssymb}
\usepackage{subfigure}

\begin{document}

\def\abs#1{ \left| #1 \right| }
\def\lg#1{ | #1 \rangle }
\def\rg#1{ \langle #1 | }
\def\lr#1{ \langle #1 \rangle }
\def\lrg#1#2#3{ \langle #1 | #2 | #3 \rangle }

\newcommand{\ud}{\overline}

\newcommand{\bra}[1]{\left\langle #1 \right\vert}
\newcommand{\ket}[1]{\left\vert #1 \right\rangle}
\newcommand{\bx}{\begin{matrix}}
\newcommand{\ex}{\end{matrix}}
\newcommand{\be}{\begin{eqnarray}}
\newcommand{\ee}{\end{eqnarray}}
\newcommand{\nn}{\nonumber \\}
\newcommand{\no}{\nonumber}
\newcommand{\de}{\delta}
\newcommand{\Del}{\Delta}
\newcommand{\lt}{\left\{}
\newcommand{\rt}{\right\}}
\newcommand{\lx}{\left(}
\newcommand{\rx}{\right)}
\newcommand{\lz}{\left[}
\newcommand{\rz}{\right]}
\newcommand{\inx}{\int d^4 x}
\newcommand{\pu}{\partial_{\mu}}
\newcommand{\pv}{\partial_{\nu}}
\newcommand{\au}{A_{\mu}}
\newcommand{\av}{A_{\nu}}
\newcommand{\p}{\partial}
\newcommand{\ts}{\times}
\newcommand{\ld}{\lambda}
\newcommand{\al}{\alpha}
\newcommand{\bt}{\beta}
\newcommand{\ga}{\gamma}
\newcommand{\si}{\sigma}
\newcommand{\ep}{\varepsilon}
\newcommand{\vp}{\varphi}
\newcommand{\ro}{\rho}
\newcommand{\tu}{\tau}
\newcommand{\z}{\mathrm}
\newcommand{\dg}{\dagger}
\newcommand{\og}{\omega}
\newcommand{\Ld}{\Lambda}
\newcommand{\m}{\mathcal}
\newcommand{\bb}{\mathbf}
\newcommand{\xb}{\underline}

\title{The initial condition problems of damped quantum harmonic oscillator}

\author{Yang Gao}\email{gaoyangchang@gmail.com}
\author{Qing Bin Tang} \author{Ru Min Wang}
\address{\it Department of Physics, Xinyang Normal University,
Xinyang, Henan 464000, China}

\begin{abstract}{We investigate the exact dynamics of the damped quantum
harmonic oscillator under the (un)correlated initial conditions. The
master equation is generalized to the cases of the arbitrary
factorized state and/or Gaussian state. We show that the variances
of the factorized Gaussian state do not sensitively depend on the
initial oscillator-bath correlation, which however can remarkably
affect the mean values even at high temperature. We also illustrate
that the correlations among the factorized states still give rise to
the initial dips during the purity evolutions, which can be smoothed
out by increasing the amount of correlation to some extent. We
finally study the effects of repeated measurements on the time
evolution of the damped oscillator analytically, which are compared
with the weak coupling results to indicate that they give rather
different transient behaviors even for an intermediate coupling.}
\end{abstract}
\maketitle

\section{Introduction}

The recent experimental developments in the field of ultrafast and
ultrasmall devices at low temperature are strongly demanding for the
fully treatments of the non-Markovian dynamics of open quantum
systems \cite{open}, which go beyond the traditional Markovian
approximation omitting the memory effects of surrounding bath in the
weak coupling limit. The rigorous analysis has been explored
extensively in the literature. For example, the exact master
equation of quantum Brownian motion was derived by the path integral
\cite{phpz,hpz,kg}, which has been extended to other open systems
\cite{ext}, such as quantum dots, the nano-cavities, and quantum
transportation in photonic crystals. On the other hand, in most of
these studies, the initial system-bath correlations are often
neglected for mathematical simplicity, the roles of which in
realistic could become significant for the strongly coupled
system-bath interactions. It is shown that the initial qubit-bath
correlations can break the completely positive property of the
evolution maps--leading to nontrivial differences in quantum
tomography process \cite{qt}, and can be witnessed by the increase
of the distance of two system states over its initial value
\cite{measure}. Besides, it is found that the effects of the initial
correlations on the photon squeezing in a cavity, the coherence of a
qubit, the entanglement of two-qubit, and etc, can induce the
oscillating dynamics in the strong non-Markovian regime
\cite{initial}.

In this paper, we continue to investigate the exact dynamics in the
presence of the initial correlations with the dissipative quantum
harmonic oscillator as an example. The initial oscillator-bath
correlations are incorporated in two different ways: (i) prepare an
initial state by implementing some measurements on an prior state,
such as a projective measurement acting only on the oscillator,
which does not change its equation of motion \cite{dist}; (ii) alter
the equation of motion of the oscillator by adjusting its
parameters--mass or frequency at an initial time, whereas the
initial state is untouched \cite{cl}. From the general solutions to
quantum Langevin equation (QLE), the time evolutions of oscillator
are simply derived by using the Wigner representations of operators
and the Gaussian properties of the total system, which allow us to
examine the existence of the master equation for the oscillator and
specify the conditions resulting in some certain master equations
straightforwardly. We also get a time-local QLE by introducing an
effective fluctuation force, and apply the modified canonical method
used by Unruh and Zurek in \cite{uz} to derive the exact master
equations for the factorized initial conditions, which can in
further include the cases of initial correlated Gaussian states.

We show that for the factorized Gaussian state, the initial
oscillator-bath correlation plays unimportant roles for the
variances over a wide range (except the regime of under-damping) of
coupling strength. However, the effect of initial correlation
becomes remarkable for the expectation values even at high
temperature. We illustrate that the initial correlations in the
factorized states are not enough to smooth out the initial dips
displayed during the purity evolutions. By increasing the amount of
initial correlations to some extent, these dips just disappear. We
also study the effects of repeated measurements on the time
evolution of the damped oscillator analytically. The comparison with
the weak coupling results is made to indicate that even for an
intermediate coupling, they have rather different transient
behaviors.

The rest of this paper is as follows. In Sec. II, we briefly review
the basics of QLE for subsequent discussions. The exact dynamics
with the initial correlations is obtained through the Wigner
representation in Sec. III. Next, in Sec. IV we use the canonical
method to derive the master equations for the factorized initial
conditions. Then in Sec. V we give some examples to show the
implications of the previous results. Finally, a short summary is
given in Sec. VI.

\section{The equations of motion}

The interacting oscillator-bath Hamiltonian to model the damped
harmonic oscillator is usually taken as \cite{qle}, \be H = {p^2
\over 2m} + { 1 \over 2 }k x^2 + \sum_j \lz {p_j^2 \over 2m_j}+ { 1
\over 2 } m_j \omega_j^2 (q_j-x)^2 \rz. \ee In the Heisenberg
picture with the natural units $\hbar=k_B=m=1$, the equations of
motion of the oscillator take the form of an initial value problem,
\be\ddot x(t)+\int _{0} ^t dt'\mu(t-t')\dot x(t')+k x(t) &=&
-\mu(t)x(0) + F(t), \nn  \dot x(t)&=& p(t),\label{qle} \ee which are
known as the QLE for dissipative harmonic oscillator. Here the
memory kernel and fluctuating force
are \be \mu(t) = \sum_j m_j\og_j^2 \cos(\og_j t), \quad t>0, \hspace{1cm} \\
F(t) = \sum_j m_j\og_j^2 \lz q_j(0) \cos(\og_j t)+
p_j(0)\frac{\sin(\og_j t)}{m_j \og_j} \rz. \ee The two quantities
are connected through the commutator, $[F(t),F(0)]=i\dot \mu(t)$,
which is necessary for the conservation of the elementary commute
relation $[x(t),p(t)]=i$ for $t\ge0$. It can be seen that $[x(0),
F(t)]=0$ and $[\dot x(0), F(t)]=0$ because $F(t)$ only depends on
the bath variables at $t=0$. The general solution of (\ref{qle})
yields to be \be x(t)&=& \dot G(t)x(0)+ G(t)p(0)+X(t), \label{sol}
\\ p(t)&=& \ddot G(t) x(0) + \dot G(t) p(0)+\dot X(t), \label{sol1}
\ee where \be X(t)=\int_0^tdt' G(t-t')F(t'), \ee and the retarded
Green function $G(t)$ for $t>0$ satisfies the equation \be \ddot
G(t)+\int _{0}^tdt' \mu(t-t')\dot G (t')+ k G(t) = 0. \label{green}
\ee and $G(t)=0$ when $t<0$. At $t=0$, the conditions $G(0)=0$ and
$\dot G(0)=1$ are imposed. In the following we set $t\ge 0$ for
convenience. Explicitly, $G(t)$ can be expressed by the Fourier
integral, \be G(t)&=&{1 \over 2\pi}\int_{-\infty}^{\infty} d\omega
\al(\omega) e^{-i\omega t}\nn &=&{2 \over \pi}\int_{0}^{\infty}
d\omega \al''(\omega) \sin(\omega t), \label{gt}\ee where the
susceptibility $\al(\og)=\al'(\og)+i\al''(\og)$ is \be
\al(\og)=\frac{1}{-\og^2-i\og\eta(\og)+k}. \ee Here \be
\eta(\og)=\eta'(\og)+i\eta''(\og)=\int_{0}^{\infty} dt \mu(t)
e^{i\omega t} \ee is the Fourier transform of the memory kernel
$\mu(t)$, of which the general properties are summarized in
\cite{qle}.

On the other hand, eliminating the dependence on the initial value
$x(0)$ and $p(0)$ of (\ref{sol}) and (\ref{sol1}) yields the local
form equation, \be \ddot x(t) + \Gamma(t) \dot x(t)+K(t) x(t) = \m
F(t), \label{local} \ee where the respective coefficients and
effective force are given by \be \Gamma(t) = {G G^{(3)}-\dot G \ddot
G\over \dot G^2- G \ddot G}, \quad K(t) = {\ddot G^2 - \dot G
G^{(3)}\over \dot G^2- G \ddot G}, \nn \m F(t) = \ddot X(t) +
\Gamma(t) \dot X(t) + K(t) X(t). \hspace{.8cm} \ee It would serve as
an elementary equation to obtain the master equations in section IV.

Usually, the initial state is taken as the uncorrelated form
\cite{hpz,solut}, \be \rho(0)=\rho_S(0) \otimes \rho_B^T, \quad
\rho_B^T={e^{-H_B/T}\over \mathrm{Tr}_B[e^{-H_B/T}]}. \ee However,
this assumption becomes problematic when the system-bath coupling
gets strong \cite{ext}. A more physical alternative is that we
prepare an initial state by performing measurements on a certain
starting state \cite{dist}, which is usually take as the time
invariant Gibbs state of the total system, \be \rho_T={e^{-H/T}
\over \mathrm{Tr}[e^{-H/T}]}.\ee In such a case $x(t)$ is a Gaussian
variable which is fully characterized by the first and second
moments. It is obvious that $\lr {x(t)}=0$, and the second moment
can be obtained through the commute relation \be [x(t),x(0)]=-i
G(t), \ee and the fluctuation-dissipation relation \cite{fd}, \be
S(t)&=& {1\over 2}\lr{x(t)x(0)+x(0)x(t)} \nn
&=&\frac{1}{\pi}\int_0^\infty d\omega \al''(\omega) \coth{\omega
\over 2 T}\cos (\omega t), \label{st}\ee where $S(t)$ is the
symmetrized correlation function, and the angular brackets denote
the expectation over the Gibbs state $\rho_T$ without further
specification. In particular, we have $\lr{x^2}=S(0)$ and
$\lr{p^2}=-\ddot S(0)$.

Instead of preparing an initial state by some measurements on the
system, where its equation of motion remains, alternatively suppose
a situation without any measurement performed while the equation of
motion is changed by adjusting some parameters of oscillator, e.g.
mass or oscillating frequency. We can get a new evolve state from
such an arrangement, which was discussed earlier by the path
integrals \cite{cl}. However, it becomes much neater with the QLE,
and the final results can be obtained by fewer steps, because all
the dynamical influences from the bath on the system are
characterized by a single memory kernel in the equation of motion.

If the changes of spring constant $k \to \z{k}$ and mass $m\to
\z{m}$ are made at $t=0$, the new QLE then becomes \be \z{m} \ddot
{\z x}_t+\int _{0} ^t dt' \mu(t-t')\dot {\z x}_{t'}+ \z k \z x_t =
-\mu(t)\z x_0+F(t), \label{grequa} \ee and the initial conditions
are $ \z x_0 = x_0$, $\z p_0=p_0$. Here the convention $\m O_t\equiv
\m O(t)$ for any operator $\m O$ is introduced. Performing the
integration by parts with the identity (\ref{green}), the general
solution of (\ref{grequa}) in terms of the new Green function $\z
G(t)$ is \be \z x_t &=& \z{m} \dot {\z G}(t) \z x_0+\z{m} {\z
G}(t)\z p_0+ \int_0^t dt'{\z G}(t-t') F(t') \nn &=& \z{m} \dot {\z
G}(t) x_0+\z{m} {\z G}(t)p_0 + \int_0^t dt'\z G(t-t') \nn && \times
\big[ m \ddot x_{t'}+\int _{0}^{t'}dt'' \mu(t'-t'')\dot x_{t''}+k
x_{t'} + \mu(t')x_0 \big] \nn &=& \frac{m}{\z{m}}x_t -\epsilon_m
[\dot {\z G}(t)x_0 + \z G(t)p_0] \nn && +\int _{0}^t dt'
[\epsilon_m\ddot {\z G}(t-t')+\epsilon_k \z G(t-t')] x_{t'},
\label{main} \ee where $\epsilon_m\equiv(m-\z{m})$ and $\epsilon_k
\equiv k-\z k$. For $\z{m}=m$ and $\z k=k$, we have the trivial
result $\z{x}_t=x_t$. In the following we restrict to the case of
$\z{m}=m=1$ for simplicity, so (\ref{main}) becomes \be \z{x}_t &=&
x_t + \epsilon_k\int _{0}^t dt' \z{G}(t-t') x_{t'}. \label{solut}
\ee

In the following discussions, the ultraviolet finite model of ohmic
dissipation we choose is the exponential cut-off, namely
$\eta'(\og)=\eta e^{-{\og^2/ \Ld^2}}$, where $\eta$ characterizes
the strength of oscillator-bath coupling. Because $\eta(\og)$ is an
analytic function in the upper half $\og$-plane as required by the
causality $\mu(t)=0$ as $t<0$, its imaginary part is connected to
the real part by the Hilbert transform. In terms of the principal
value integral and the imaginary error function, we have \be
\eta''(\og)=\frac{1}{\pi}\m P\int_{-\infty}^\infty dz
\frac{\eta'(z)}{\og-z} = \eta e^{-{\og^2/ \Ld^2}} \mathrm{erfi}({\og
\over \Ld}), \ee which allows us to obtain the expressions of $G(t)$
and $S(t)$ by substituting $\eta(\og)$ into (\ref{gt}) and
(\ref{st}) respectively.

\section{Exact time evolutions for the initial correlated states}

Now let us first study the time evolution of a class of initial
correlated states prepared by a measurement $f_0=f(x_0,p_0)$ on the
Gibbs state $\rho_T$ at time $t=0$, which results in \be \rho_T \to
\rho_0={1 \over Z}{f_0\rho_T f_0^\dg}, \quad Z=\lr{f_0^\dg
f_0}=\mathrm{Tr}[f_0^\dg f_0\rho_T]. \ee It is found that the final
state at time $t>0$ can be represented by the Wigner characteristic
function with much convenience, which is defined as \be
\widetilde{W}(P,Q,t)=\mathrm{Tr}[e^A \rho_0]={1 \over Z}\lr
{f^\dg_0e^{A}f_0} \ee with the operator $A=-i[x(t)P+p(t)Q]$. Next,
we express the arbitrary operator $f$ by its Wigner representation
\be f(x,p) &=& \frac{1}{2\pi}\int d\Sigma' \widetilde{f}_w(P',Q')
e^{i(P'x+Q'p)}, \nn \widetilde{f}_w(P',Q') &=& \frac{1}{2\pi}\int
d\si' f_w(x',p')e^{-i(P'x'+Q'p')}\nn
&=&\mathrm{Tr}[f(x,p)e^{-i(P'x+Q'p)}], \ee where $d\Sigma=dPdQ$ and
$d\si=dxdp$. For the case of the density matrix $\rho$,
$\rho_w/(2\pi)$ and $ \widetilde{\rho}_w$ are the usual Wigner
distribution function $W(x',p')$ and characteristic function
$\widetilde{W}(P',Q')$. Hence the Wigner characteristic function can
be obtained by \be \lr {f^\dg_0e^{A}f_0}&=& \frac{1}{(2\pi)^2}\int
d\Sigma' d\Sigma'' \widetilde{f}_w^*(P',Q')\widetilde{f}_w(P'',Q'')
\nn && \times \lr {e^{-i(P'x_0+ Q'p_0)}e^{A}e^{i(P''x_0+Q''p_0)}},
\label{wig} \ee where the quantity in the angular bracket is found
to be $e^{-\Gamma}$ with the help of the identity \be \lr {e^{\m
O}}=\exp\lx {1\over 2}\lr {\de \m O^2} +\lr {\m O}\rx, \quad \de \m
O=\m O-\lr {\m O} \ee  for arbitrary Gaussian state and Gaussian
variable $\m O$, \be \Gamma&=& [i (\m G P_c -\dot{\m G} Q_c) +\m S
P_r- \dot {\m S} Q_r] +{1 \over 2}\big[i(Q_r P_c-P_r Q_c) \nn &&
+\lr {x^2} P_r^2 + \lr {p^2} Q_r^2+\lr{x^2}P^2+\lr{p^2}Q^2
\big].\label{wig1} \ee The introduced symbols are $\m G=GP+\dot G
Q$, $\m S=SP+\dot S Q $, $P_c={(P'+P'') / 2}$, $Q_c={(Q'+Q'')/2}$, $
P_r={P'-P''}$, and $Q_r={Q'-Q''}$. The final expression (\ref{wig1})
gives a quite simpler form of the exact time evolution for the
dissipative oscillator compared to Eq.(13) in \cite{kg}, which would
facilitate the discussions on the possible existence of the master
equation associated with $\widetilde{W}(P,Q,t)$.

Transform $\widetilde{f}_w$ back to $f_w$, we have the equivalent
expression \be \lr {f^\dg_0e^{A}f_0} = \frac{1}{\pi^2}\int
d\si'd\si'' f_w^*(x', p')f_w(x'',p'') e^{-\Gamma} \ee in terms of
the transformed coordinates $x_c=({x}'+x'')/2$, $p_c=(p'+p'')/2$,
$x_r={x}'-x''$, and $p_r={p}'-p''$, \be \Gamma &=& 2\big[i( \m G
p_c+ \dot {\m G}x_c)-(\m S-2\lr {x^2}\dot {\m G} )p_r-(\dot {\m
S}+2\lr {p^2}\m G )x_r \nn && - i (x_r p_c-p_r x_c )+\lr {x^2} p_r^2
+ \lr {p^2}x_r^2 +\m G \dot {\m S} -\dot {\m G} \m S \nn && + \lr
{x^2}\dot {\m G}^2+ \lr {p^2}\m G^2 \big]+{1\over
2}(\lr{x^2}P^2+\lr{p^2}Q^2). \ee If the preparation functions only
depend on the position, the above equation gives the result
\cite{initial}, \be \langle f^\dg_0 e^{A} f_0 \rangle
&=&\frac{1}{\sqrt{2 \pi \lr {x^2}}}\int dx' f^*(x'+{\m G \over 2}) f
(x'-{\m G \over 2})e^{-\Gamma}, \nn \Gamma &=&{x'^2 +2i x'\m S-\m
S^2 \over 2\lr {x^2}}+\frac{\lr{x^2}P^2+\lr{p^2}Q^2}{2}. \label{fx}
\ee

The method used here can be directly generalized to more complicated
cases, such as interrupting the system by multiple measurements
$f_j$ at different times $t_j$, $j=0,1,\dots,n$, which gives the
joint probability of finding the system at the states represented by
$f_j$ at $t_j$, $\mathrm{Pr}(n,\dots,1,0)=\lr {f_0^\dg f_1^\dg \dots
f_n^\dg f_n \dots f_1 f_0}$. The evaluation of this quantity is the
same as above but with many simple and lengthy expressions, so we
will confine to the case of $n=2$ in section V to consider the
effects of multiple measurements and dissipation on the system
evolution.

Next, we consider the time evolution of oscillator under a sudden
change of spring constant at $t=0$ without any measurement
performed. In such a case, we find the new Wigner characteristic
function using the solution given by (\ref{solut}), \be &&\widetilde
W(P,Q,t) = \lr{e^{-i [\z x_tP+\z p_tQ]}} = e^{-( \si_{x}P^2+\dot
\si_{x} PQ+\si_{p}Q^2)/2}, \nn &&\si_{x} = \lr{\z x^2_t} = \lr {x^2}
+ 2 \epsilon_k \int_0^tdt' \z{G}(t')S(t')\nn && \hspace{2cm} +
\epsilon_k^2 \int_0^t \!\!\!\!\!\!\ \int_0^t dt'dt''
\z{G}(t')\z{G}(t'')S(t'-t''), \nn &&\si_{p} = \lr{\z p^2_t}=\lr{p^2}
+ 2 \epsilon_k\int_0^tdt' \dot{\z{G}}(t')\dot S(t')\nn &&
\hspace{2cm} + \epsilon_k^2\int_0^t \!\!\!\!\!\!\ \int_0^t
dt'dt''\dot {\z{G}}(t') \dot{\z{G}}(t'')S(t'-t''). \label{second}
\ee For $t=0$, it in deed reduces to the expected initial state \be
\widetilde W(P,Q,0)=\exp\lx -\frac{\lr{x^2}P^2+\lr{p^2} Q^2}{2}\rx.
\label{init}\ee

At last, we consider the possible reduced master equation
\cite{kg,rp} for the oscillator under the time evolution described
by (\ref{wig}) and (\ref{wig1}) with an arbitrarily chosen
preparation measurement $f(x,p)$. That is to find an equation for
$\widetilde{W}$ connecting its time and $P,Q$ derivatives. Because
the quadratic structure of the exponent $\Gamma$, and its dependence
on the four coordinates $P_c,Q_c,P_r,Q_r$, we only need to compute
the first order derivatives over $t$, $P$ and $Q$ to check if the
following equation exists, \be {\p \widetilde{W}\over \p t}=\lx
X_1(t){\p \over \p Q}+X_2(t){\p \over \p P}+X_3(t) \rx
\widetilde{W}, \label{mass} \ee where the post-determined
coefficients $X_j(t)$ only depend on $P,Q$. Simple considerations
reveal that there is usually no solution to (\ref{mass}), since the
time derivative gives four independent terms inside the integrals,
which linearly depend on $P_c,Q_c,P_r,Q_r$, and can not be written
as the linear superposition of two independent terms from the $P,Q$
derivatives in general. However, there are some exceptions: (a)
Several coordinates can be integrated out explicitly for the
particular form of $f(x,p)$, such as $f=f(x)$ in (\ref{fx}), where
we can integrate out the two momentum coordinates and find the
master equation takes the form of (\ref{mass}) with the following
coefficients \cite{kg}, \be X_1= P-\ga_1 Q, \quad X_2=-\ga_2 Q,
\quad X_3= -D_1 PQ-D_2 Q^2, \no \ee and \be \ga_1 &=& {G \ddot
S-\ddot G S\over \dot G S-G \dot S}, \quad \ga_2={\dot G \ddot
S-\ddot G \dot S\over \dot G S-G \dot S}, \nn
D_1&=&\ga_2\lr{x^2}-\lr {p^2}, \quad D_2=\ga_1 \lr{p^2}. \ee (b)
Instead of obtaining a master equation independent of the
preparation measurements, it is possible to find one depending on
the preparation in some circumstances. (c) For the factorized
initial states, we can always find the corresponding master
equations as shown below.

\section{Master equations for the factorized initial states}

As the mostly adopted initial state, we consider the time evolution
of the uncorrelated initial system-bath state, $\rho_0=\rho_S
\otimes \rho_B^T$ with the independent system state $\rho_S$ and the
Gibbs state of bath $\rho_B^T$ to compare with the previous results.
The Wigner characteristic function then follows as
\cite{solut,local}, \be \widetilde W(t)&=& \lr {e^{A}}= \mathrm {Tr}
\lz e^{A} \rho_S \otimes \rho_B \rz\\ &=& \mathrm {Tr}_S \lz
e^{-i[(\dot G x_0+ G p_0)P+(\ddot G x_0 + \dot G p_0)Q]} \rho_S \rz
\nn &&\times \mathrm {Tr}_B \lz e^{-i (X P + \dot X Q)} \rho_B \rz
\nn &=& \widetilde W(\dot {\m G},\m G,0) \exp \lz -{1\over 2} \lx
b_{x} P^2+\dot b_{x}PQ+ b_{p} Q^2\rx \rz, \no \ee where we have used
(\ref{sol})-(\ref{sol1}) and the facts that $x_0,p_0$ and $X(t)$ for
$t \ge 0$ are commutable, and that $X,\dot X$ are Gaussian
variables. Specifically, the angular brackets in this section denote
the average over the initial state $\rho_0$. The moments appearing
in the above equation can be written as \be b_{x} \equiv \lr{X^2}
&=& \int_0^\infty d\og E_T(\og) \abs {\int_0^t dt' G(t') e^{i\og
t'}}^2, \nn b_{p} \equiv \lr{\dot X^2} &=& \int_0^\infty d\og
E_T(\og) \abs {\int_0^t dt' \dot G(t') e^{i\og t'}}^2, \nn E_T(\og)
&=& \eta'(\og) \frac{\og}{\pi}\coth{\og \over 2 T}. \ee


The master equation can be easily derived by the method of the last
section, since we have only two free coordinates here. However, we
will use the canonical method introduced by Unruh and Zurek in
\cite{uz} with a slight modification to re-derive the master
equation for the possible generalization later on. According to
\cite{uz}, the time derivative of the Wigner characteristic function
can be evaluated by the identity \be \frac{d e^A}{dt} = \int_0^1
d\ld e^{(1-\ld) A} \dot A e^{\ld A}, \ee which becomes obvious from
another identity \be e^{A+B}=e^A\bigg[ 1+\int_0^1 d\ld e^{-\ld A} B
e^{\ld (A+B)}\bigg]. \ee Hence we get \be && \frac{\p\widetilde
W}{\p t} = \int_0^1 d\ld \lr {e^{(1-\ld) A} \lz -i (\dot x P+\ddot x
Q)\rz e^{\ld A}} \nn &&= \int_0^1 d\ld \lr {e^{(1-\ld) A} \lt -i \lz
\dot x P+\lx \m F-\Gamma\dot x-K x\rx Q \rz \rt e^{\ld A}} \nn &&=
(P-\Gamma Q) \frac{\p \widetilde{W}}{\p Q}-K Q\frac{\p
\widetilde{W}}{\p P} \nn && \indent + \int_0^1 d\ld \lr {e^{(1-\ld)
A} \lx -i Q\m F\rx e^{\ld A}}, \label{master} \ee where the local
equation (\ref{local}) has been used to reduce the second order
derivative and the last integral can be evaluated through \be
\frac{\p}{\p \alpha}\lr {e^{A-i Q\m F \alpha}}\big |_{\alpha=0} =
-(D_x PQ+D_p Q^2)\widetilde{W} \ee with $D_x=\lr {\m F X +X\m F}/2$
and $D_p=\lr {\m F\dot X +\dot X \m F}/2$. It thus yields the final
result \be \frac{\p\widetilde W}{\p t}=\lx P \frac{\p}{\p Q}-\Gamma
Q \frac{\p}{\p Q}-K Q\frac{\p}{\p P} - D_x PQ -D_p Q^2 \rx
\widetilde W.\no \ee The explicit expressions for $D_x$ and $D_p$ by
definitions are \be D_x &=& {1\over 2}\ddot b_{x}+{\Gamma\over
2}\dot b_{x}+{K}b_{x}-b_{p} \nn D_p &=& {1\over 2}\dot b_{p}+{K\over
2}\dot b_{x}+{\Gamma}b_{p}. \ee

These two coefficients are the same as Eq. (89) in \cite{kg}, but
different from Eq. (3.8) in \cite{solut}, where the postulation $\m
F(t)=F(t)$ was implicitly assumed. In fact, It is not legetimate for
the cases with the non-local memory kernels \cite{local}, i.e.
$\mu(t)\neq \eta\delta(t)$. However, they can still be viewed as the
weak coupling approximations for the exact results, and provide a
simpler expression of the master equation in the weak coupling
limit. Particularly, we have the approximated coefficients up to
$O(\eta)$, \be \Gamma = -\int_0^t\!\!\!\!\ dt' G_0(t') \dot \mu(t'),
\quad K = k+\int_0^t\!\!\!\!\ dt'\dot G_0(t') \dot \mu(t'), \nn D_x
= \int_0^t\!\!\!\!\ dt' G_0(t')\nu(t'), \quad D_p =
\int_0^t\!\!\!\!\ dt' \dot G_0(t')\nu(t'),\hspace{.5cm} \ee where
$G_0(t)={\sin (\og_0 t)/\og_0}$ with $\og_0=\sqrt k$ is the Green
function without dissipation and the correlation function
$\nu(t)=\lr{F(t)F(0)+F(0)F(t)}/2=\int_0^\infty d\og E_T(\og)
\cos(\og t)$.

It also needs to point out that the conclusions drawed from equation
(A9) in \cite{hpz}--especially the exact master equations with the
uncorrelated initial states still have serious divergence problems
due to the zero point energy even in the ultraviolet cut-off
models--are misleading, where the authors unconsciously omit a term
proportional to $F(0)$ from (A7) to (A8)--because $\dot G(0)=1$ and
$\ddot x_s(t)=F(t)+\int_{-\infty}^t dt'\ddot G(t-t')F(t')$, which
would cancel the divergence displayed in (A9).

The master equation for the Wigner function can be obtained by the
replacements \be P \leftrightarrow -i \frac{\p}{\p x}, \quad
\frac{\p}{\p P} \leftrightarrow -ix, \quad Q \leftrightarrow -i
\frac{\p}{\p p}, \quad \frac{\p}{\p Q} \leftrightarrow -ip,\no \ee
and thus \be \frac{\p W}{\p t}&=&\bigg( -\frac{\p}{\p x}p+\Gamma
\frac{\p}{\p p}p+K \frac{\p}{\p p}x \nn && + D_x \frac{\p^2}{\p x \p
p}+D_p \frac{\p^2}{\p p^2} \bigg) W. \label{master1} \ee
Furthermore, we can get the master equation for the density matrix
$\rho_S$ by the replacements \be [x,\cdot] \leftrightarrow
i\frac{\p}{\p p}, \quad \lt x,\cdot \rt {\leftrightarrow} 2x, \quad
\lz p, \cdot \rz \leftrightarrow -i\frac{\p}{\p x},\quad \{p,\cdot\}
\leftrightarrow 2p, \no \ee which leads to \cite{hpz} \be \frac{\p
\rho_S}{\p t}&=&-i[H_R,\rho_S]-{i \over 2} \Gamma [x,\{p,\rho_S\}]
\nn && + D_x [x,[p,\rho_S]]-D_p [x,[x,\rho_S]]. \ee Here
$H_R={p^2/2}+K(t) x^2/2$ is the renormalized Hamiltonian of the
system alone.

On the other hand, for the factorized states where the bath takes an
oscillator-dependent state $\rho_B(\rho_S)$, instead of the
independent Gibbs state, we still have
$\widetilde{W}(P,Q,t)=\widetilde{W}(\m {\dot G},\m G,0)B(P,Q,t)$
with $B=\mathrm{Tr}[e^{-i(PX+Q\dot X)}\rho_B]$. Using the explicit
expression of the effective force, the last integral in
(\ref{master}) becomes \be && \int_0^1 d\ld \lr {e^{(1-\ld) A} \lz
-i Q\m (\ddot X+\Gamma \dot X+K X)\rz e^{\ld A}} \nn &&=
\widetilde{W}(\m {\dot G},\m G,0)\lz {\p \over \p t} + (-P+\Gamma Q)
{\p \over \p Q} + K Q {\p \over \p P} \rz B \nn && \equiv
\widetilde{W}(\m {\dot G},\m G,0) \m D[B]. \ee It yields the master
equation of the form \be \m D [\ln \widetilde{W}]=\m D[\ln B], \ee
which can be obtained more straightforwardly by reminding the
identity $\m D[\ln \widetilde{W}(\m {\dot G},\m G,0)]=0$. Hence
(\ref{wigf}) becomes evident where $\ln B=-(b_x P^2+\dot b_x PQ+b_p
Q^2)/2$.

Moreover, for an arbitrary initial correlated Gaussian state
$\rho_0$ being the quadratic function of dynamical variables, we
have \be \widetilde W(P,Q,t) &=& \exp \bigg[ -\frac{1}{2}(\si_{x}
P^2+\dot \si_{x}PQ+ \si_{p}Q^2) \nn && -i(\lr{x_t}P+\lr{p_t}Q)
\bigg], \label{gauss} \ee where $\si_{x}=\lr{\delta x_t^2}$, $\delta
x_t=x_t-\lr{x_t}$, and etc. It also allows for a master equation.
Repeat the steps to (\ref{master}), where the last integral is
evaluated to be $ (-iQ\lr {\m F}-D_x PQ-D_p Q^2)\widetilde W$ with
two different coefficients \be D_x&=&\frac{1}{2}\lr {\m F x +x\m
F}-\lr {\m F} \lr x,\\ D_p &=& \frac{1}{2}\lr {\m F p +p\m F}-\lr
{\m F} \lr {p}. \ee The master equation thus takes the form as \be
\frac{\p W}{\p t}&=&\bigg( -\frac{\p}{\p x}p+\Gamma \frac{\p}{\p
p}p+K \frac{\p}{\p p}x-\lr {\m F}\frac{\p}{\p p}\nn && + D_x
\frac{\p^2}{\p x \p p}+D_p \frac{\p^2}{\p p^2} \bigg) W.
\label{master2} \ee where an extra term describing the drift due to
a non-zero expectation of the effective force appears. Unlike
(\ref{master1}), there are three initial state-dependent
coefficients in (\ref{master2}), which incorporate the initial
oscillator-bath correlations. On the other hand, the time derivative
of (\ref{gauss}) can be represented by \be \frac{\p\widetilde W}{\p
t} &=& \bigg[-\frac{1}{2}\lx \dot \si_{x} P^2+\ddot \si_{x}P Q +\dot
\si_{p} Q^2 \rx \nn && -i(\lr{\dot x}P+\lr{\dot p}Q) \bigg]
\widetilde W, \ee which is another equivalent form of the above
master equation. The reason is that the density matrix is over
determined in this situation. Transforming to the Wigner function,
we have \be \frac{\p W}{\p t}&=&\bigg[ \frac{1}{2}\lx \dot \si_{x}
{\p^2 \over \p x^2}+\ddot \si_{x}{\p^2 \over \p x \p p} +\dot
\si_{p} {\p^2 \over \p p^2} \rx \nn && -\lr {\dot x}{\p \over \p
x}-\lr {\dot p}{\p \over \p p} \bigg] W, \label{gauss}\ee which is
similar to the classical diffusion equation with time varying
coefficients. However, the simplicity of (\ref{gauss}) is sacrificed
by introducing more initial state-dependent coefficients.

\section{Examples}

\begin{figure}[t!]\label{f1}
 \centering \subfigure[]{ \label{Fig.sub.a}
\includegraphics[width=0.45\columnwidth]{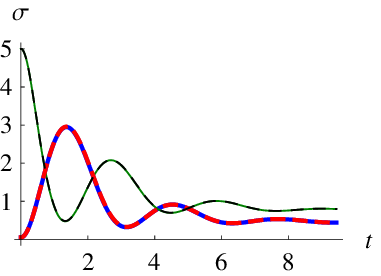}}\hspace{0.05in}
 \subfigure[]{\label{Fig.sub.b}
\includegraphics[width=0.4\columnwidth]{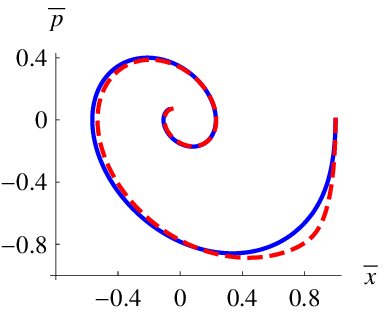}}\hspace{0.05in}
\caption{(a) The time evolutions of the (thick) position and (thin)
momentum variances $\si_x,\si_p$ for the correlated-(solid) and
uncorrelated-(dashed) initial conditions with the finite Ohmic bath
spectral density. (b) The $\bar x_t\bar p_t$-trajectories with the
correlated-(solid) and uncorrelated-(dashed) initial conditions in
the phase space. The parameters here are $\eta=0.5$, $k=1$,
$\ld=10$, and $x_0=1$.}
\end{figure}

In this section, we select some examples to show the utility of our
exact results. At first, suppose the projective operator \be
f(x,p)=\lg \psi \rg \psi, \quad \lg \psi = e^{-i \bar x_0 p} e^{ir
(xp+px)/2} \lg 0 \ee as the displaced squeezed state and $\lg 0$
being the vacuum for free oscillator, is applied on the oscillator.
The resulting initial state takes the factorized form, \be \rho_0 =
\lg \psi \rg \psi \otimes {\rg \psi \rho_T \lg \psi \over Z}, \quad
Z=\mathrm{Tr}_B[\rg \psi \rho_T \lg \psi]. \label{factor} \ee This
factorized state still captures the system-bath correlation by the
classical dependence of the bath state on the system. Putting \be
\widetilde{f}_w(P,Q) = \exp\lz -\frac{1}{2} \lx \de_x P^2+\de_p
Q^2\rx-i\bar x_0 P \rz, \ee associated with $\de_p = \ld /2$,
$\de_x=1/(2\ld)$, and $\ld=e^{2r}$ into (\ref{wig}) and calculating
the four-fold integrals, we obtain \be \widetilde W(P,Q,t) &=&
\exp\bigg[ -\frac{1}{2}\lx \sigma_{x} P^2+\dot \sigma_{x} PQ+
\sigma_{p} Q^2 \rx \nn && -i \lx \bar x_t P+{\bar p}_t Q \rx \bigg],
\label{wigf} \ee where \be
\sigma_{x}&=&\lr{x^2}+{\ld^2G^2+\dot{G}^2\over\ld}-{(\dot{G}+2\ld
S)^2\over 2\ld w_x}-{(\ld G-2\dot{S})^2 \over 2\ld w_p}, \nn
\sigma_{p}&=& \lr{p^2}+{\ld^2 \dot
G^2+\ddot{G}^2\over\ld}-{(\ddot{G}+2\ld\dot S)^2 \over 2\ld
w_x}-{(\ld \dot G-2\ddot{S})^2 \over 2\ld w_p}, \nn \bar x_t &=&
{\dot G+2\ld S \over w_x} \bar x_0, \quad \bar p_t=\dot{\bar x}_t,
\label{ff} \ee with $w_x={1+4\de_p \lr {x^2}}$, $ w_p={1+4\de_x\lr
{p^2}}$, and $Z=\sqrt{w_x w_p}/2$. For the uncorrelated initial
state \be\rho_0=\lg \psi \rg \psi \otimes \rho_B^T, \label{stat2}\ee
the Wigner characteristic function is also given by Eq.(\ref{wigf})
except with the different coefficients \be \sigma_{x} &=& G^2 \de_p
+\dot G^2\de_x +b_{x}, \nn \sigma_{p}&=& \dot G^2 \de_p + \ddot
G^2\de_x+b_{p}. \nn \bar x_t &=& \dot G \bar x_0,\quad {\bar p}_t
=\dot {\bar x}_t. \label{fff} \ee

\begin{figure}[t!]
\begin{minipage}{.20\textwidth}
\centerline{\epsfxsize 45mm \epsffile{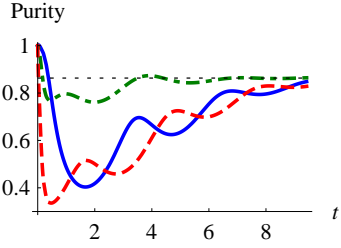}}
\end{minipage}
\caption{This plot shows the time evolutions of the purity
$\z{Tr}[\rho_S^2]$ with different squeezing parameters, $\ld=10$
(solid), $1$ (dot-dashed), $0.1$ (dashed). The other parameters are
chosen as in Fig. 1.}
\end{figure}

To see the implication of the above results, we resort to the
numerical results pictorially. The parameters chosen for numerical
computations are $\Lambda=10$ and $T=0$. In Fig. 1(a), we note  for
the factorized Gaussian states, the differences between the
correlated and uncorrelated initial conditions are nearly
unnoticeable for the evolutions of the position and momentum
variances (\ref{factor}) and (\ref{stat2}). However, the $\bar
x_t\bar p_t$-trajectories in the phase space follow quite different
pathes as displayed in Fig. 1(b). Such phenomena have also been
observed in \cite{gauss} before.  More importantly, from Eqs.
(\ref{ff}) and (\ref{fff}), this difference is still remarkable at
high temperature because the correlation function $S(t)$ depends on
temperature and for $T \gg \Lambda \gg \og_0$, we have $\lr
{x^2}\approx T/k$, $S(t)\approx {T/ k} - T \int_0^t dt' G(t')$, and
\be \bar x_t\approx \bigg[1-k\int_0^t dt' G(t')\bigg]\bar x_0 \neq
\dot G \bar x_0. \ee

Next, we consider the time evolution of the purity defined as
$\sigma_\z{purity}= \z{Tr}[\rho_S^2]$, or \be \sigma_\z{purity}=
\frac{1}{2\pi}\int d\Sigma|\widetilde{W}(P,Q,t)|^2 =
\frac{1}{\sqrt{4\si_x \si_p-\dot \si_x^2}}. \ee In Fig. 2, we see
the appearance of the initial dips at short times due to the initial
jolts \cite{hpz,uz} of the coefficients in the master equation.
Therefore, the correlations among the factorized states are not
enough to smooth out these dips. Because the purity does not
sensitively depend on the initial conditions as shown above. We can
use Eq. (\ref{fff}) to expand the purity at short times to find
\be\sigma_\z{purity} \approx 1-\frac{\Lambda^2t^2}{2\pi \ld},\ee
which sharply decreases for times $t\ll \ld^{1\over 2}/\Lambda$ as
shown in Fig. 2. For long times, it approaches to the equilibrium
value $\sigma_\z{purity}=0.8658$. If we use the second way discussed
in Sec. II to obtain an evolved state described by Eq.
(\ref{second}), the comparison of which with Eq. (\ref{fff}) and
$\rho_S(0)$ characterized by Eq. (\ref{init}) is shown in Fig. 3. It
can be seen that they give different evolutions for the variance and
purity. Particularly, we note that the initial dip associated with
the factorized state does not show up for the new state, which dues
to the correlation in the initial Gibbs state of the whole system is
stronger than in the factorized initial state to smooth out the dip.

\begin{figure}[t!]\label{f1}
 \centering \subfigure[]{ \label{Fig.sub.a}
\includegraphics[width=0.45\columnwidth]{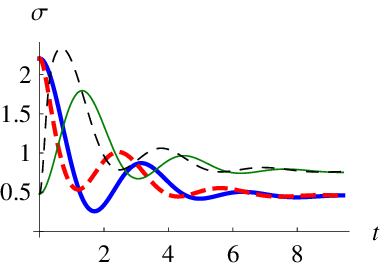}}\hspace{0.05in}
 \subfigure[]{ \label{Fig.sub.b}
\includegraphics[width=0.45\columnwidth]{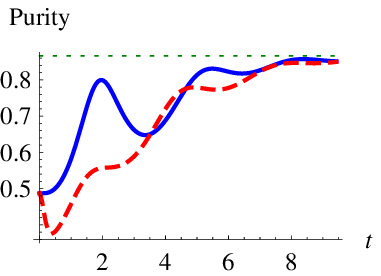}}\hspace{0.05in}
\caption{(a) The solid lines show the (thick) position and (thin)
momentum variances $\si_x,\si_p$ for the evolved state after the
sudden change of the oscillating frequency $k=0.01 \to \z k=1$ at
$t=0$. The dashed lines represent the relevant results under the
uncorrelated initial condition. (b) The purity evolutions under the
above two different conditions. Here $\sigma_\z{purity}(0)=0.4870$
and $\eta=0.5$.}
\end{figure}

\begin{figure}[t!]
\begin{minipage}{.20\textwidth}
\centerline{\epsfxsize 45mm \epsffile{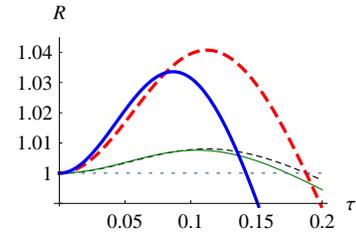}}
\end{minipage}
\caption{The plot of the survival probability ratio, $R=\z
{Pr}(2,1,0)/\z {Pr}(2,0)$ versus the time interval $\tau$, for the
exact-(solid) and approximated-(dashed) results with the (thick)
intermediate coupling constant $\eta=0.5$ and (thin) weak coupling
constant $\eta=0.1$. The spring constant is $k=1$.}
\end{figure}

Finally, we study the effects of two measurements $f_j=\lg 0_j \rg
0$ applied at times $t_j=j \tau$, $j=1,2$, during the evolution of
the initial state prepared by $f_0=\lg 0 \rg 0$ for simplicity. The
quantities of interest are the joint conditional probabilities
$\mathrm {Pr}(2,1|0)={\mathrm {Pr}(2,1,0)/\mathrm {Pr}(0)}$ and
$\mathrm {Pr}(2|0)={\mathrm {Pr}(2,0)/\mathrm {Pr}(0)}$, which can
be used to define the survival ratio \be R ={\mathrm {Pr}(2,1|0)
\over \mathrm {Pr}(2|0)}= {\lr {f_0^\dg f^\dg_1 f^\dg_2 f_2 f_1 f_0}
\over \lr {f_0^\dg f^\dg_2 f_2 f_0}}. \ee The meaning of $R>1$
($R<1$) is that the intermediate measurement at $\tau$ enhances
(suppresses) the survival probability of the initial state at
$2\tau$, while $R=1$ indicates the crossover point. The explicit
expression for $R$ involves the determination of a $10\times 10$
matrix, so it is too lengthy to put here and we only plot the
numerical results in Fig. 4. A similar problem has been previously
studied in \cite{zeno} with the weak coupling and secular
approximations to neglect the initial and subsequential
oscillator-bath correlations and the fast oscillating terms in the
master equation, respectively, and obtain a simplified equation for
$\mathrm {Pr}(n,\dots,1|0) \approx \exp[-n\ga(\tau)]$ with
$\ga(t)=\int_0^tdt' D_p(t')-{\og_0/2}\int_0^tdt' \Gamma(t')$. From
Fig. 4, we see that for the weak coupling $\eta=0.1$, the two
results are almost the same. However, for a modest coupling
$\eta=0.5$, they could become significantly different. Contrary to
the conclusion in the weak coupling limit, the crossover points
predicted by the exact results vary with the coupling constant
sensitively, and are always less than the approximated ones.


\section{Summary}

In conclusion, we took the damped quantum harmonic oscillator as an
example and applied the QLE and Wigner representation for operator
to investigate the exact dynamics in the presence of the initial
oscillator-bath correlation incorporated in two different ways: (i)
prepare an initial state by a projective measurement on the
oscillator; (ii) change the equation of motion of the oscillator by
adjusting its parameters--mass or frequency initially. The simpler
results thus obtained facilitate us to defy the possibility of the
master equation independent of the initial state and specify several
sufficient conditions resulting in some certain master equations. We
also got a time-local QLE to derive the exact master equations under
the factorized initial conditions, including the cases of initial
correlated Gaussian states. It was shown that the variances of the
factorized Gaussian states do not sensitively depend on the initial
oscillator-bath correlations, which can however significantly
influence the mean values even at high temperature. We demonstrated
that the correlations among the factorized states still give rise to
the initial dips during the purity evolutions, which can be smoothed
out by increasing the amount of initial correlation to some extent.
We finally studied the effects of repeated measurements on the
evolution of the damped oscillator, which were compared with the
weak coupling results to indicate that they give rather different
transient behaviors even for an intermediate coupling.

\begin{acknowledgments}
We would like to thank Prof. R.F. O'Connell for helpful discussions.
This work is supported by NSFC grand No. 11304265, the Education
Department of Henan Province (No. 12B140013), and the Program for
New Century Excellent Talents in University (No. NCET-12-0698).
\end{acknowledgments}

\section*{APPENDIX}

To introduce the Wigner representation of an arbitrary operator
$A(x,p)$, let us first review the normal order of $A$, where $x$ is
always put in front of $p$ in any product, denoting by $A=
:A_n(x,p):$, such as $px=:xp-i:$. In particular, we can take $x,p$
as $c$-numbers inside the symbol $:$ $:$, i.e. $:xp:=:px:$.
Inserting the unity decomposition, we have \be A(x,p) &=& \int d\si'
\lg {x'} \rg {x'} :A_n(x,p): \lg {p'} \rg {p'} \nn &=& \int d\si'
\lg {x'} \rg {x'} A_n(x',p')\lg{p'}\rg {p'}\nn &=&
\frac{1}{\sqrt{2\pi}}\int d\si' \lg {x'}\rg {p'} e^{ip'x'}
A_n(x',p'), \ee and thus
$A_n(x',p')=\sqrt{2\pi}\lrg{x'}{A(x,p)}{p'}e^{-ip'x'}$. The
composition rule $C(x,p)=A(x,p)B(x,p)$ can be transformed into \be
C(x,p)&=&\int d\si' d\si'' \lg {x'} \rg {x'} A_n(x',p')\lg {p'} \rg
{p'}\nn && \times \lg {x''} \rg {x''}B_n(x'',p'')\lg {p''} \rg {p''}
\nn &=& \frac{1}{\sqrt{2\pi}} \int dx' dp'' \lg {x'}\rg {p''}
e^{ip''x'} C_n(x',p''), \no \ee where \be C_n(x',p'')&=&
\frac{1}{2\pi}\int dp'dx'' A_n(x',p') B_n(x'',p'') \nn && \times
e^{i(p'-p'')(x'-x'')}. \ee

On the other hand, we have \be A(x,p) &=& \int d\si'  A_n(x',p')
\delta(x'-x)\delta(p'-p) \nn &=& \frac{1}{(2\pi)^2}\int
d\si'd\Sigma' A_n(x',p') e^{iP'(x-x')}e^{iQ'(p-p')} \nn &=&
\frac{1}{2\pi}\int d\Sigma' \widetilde{A}_n(P',Q')
e^{i(P'x+Q'p)-iP'Q'/2}, \label{gw} \ee where we defined the normal
order characteristic function \be
\widetilde{A}_n(P',Q')&=&\frac{1}{2\pi}\int d\si' A_n(x',p')
e^{-i(P'x'+Q'p')} \nn &=& \z {Tr}[A e^{-i(P'x+Q'p)+iP'Q'/2}].
\label{gw1}\ee The appearances of $-i P'Q'/2$ in the last
exponentials of (\ref{gw}) and (\ref{gw1}) dues to the
non-commutability of $x,p$. If we insist on omitting this term, the
above equation should be re-interpreted as the Wigner representation
or symmetrized order of the corresponding operator rather than the
normal order, namely $A_n(x',p')\to A_w(x',p')$. In fact, they can
be further generalized to \be A(x,p) &=& \frac{1}{2\pi}\int d\Sigma'
\widetilde{A}_g(P',Q')e^{i(P'x+Q'p)-igP'Q'/2}, \nn
\widetilde{A}_g(P',Q')&=&\frac{1}{2\pi}\int d\si' A_g(x',p')
e^{-i(P'x'+Q'p')} \nn &=& \z {Tr}[A e^{-i(P'x+Q'p)+igP'Q'/2}], \ee
which includes the usual cases of $g=0$ or $\pm 1$ as the Wigner or
(anti)-normal order representations.

As an example, take $A(x,p)=\lg 0 \rg 0$, then $\rg {x'} 0\rangle =
{e^{-x'^2/ 2} / {\pi}^{1\over 4}}$, $\rg {0} p'\rangle = {e^{-p'^2/
2}/{\pi}^{1\over 4}}$, and \be A_n=\sqrt{2\pi} \rg {x'} 0\rangle \rg
0 {p'}\rangle e^{-ip'x'}=\sqrt{2}e^{ -(x'^2+2ix' p'+p'^2)/2}, \nn
\widetilde{A}_n=e^{ -(P'^2+2iP' Q'+Q'^2) /4}, \quad
\widetilde{A}_w=e^{ -(P'^2+Q'^2)/4}. \hspace{.6cm} \ee

\end{document}